# Metastable and Unstable Cellular Solidification of Colloidal Suspensions

Sylvain Deville[1*], Eric Maire[2], Guillaume Bernard-Granger[1], Audrey Lasalle[1], Agnès Bogner[2], Catherine Gauthier[2], Jérôme Leloup[1], Christian Guizard[1]

[1] Laboratoire de Synthèse et Fonctionnalisation des Céramiques, UMR3080 CNRS/Saint-Gobain, Cavaillon, France

[2] Université de Lyon, INSA-Lyon, MATEIS CNRS UMR5510, Villeurbanne, France

**Abstract**

Colloidal particles are often seen as big atoms that can be directly observed in real space. They are therefore playing an increasingly important role as model systems to study processes of interest in condensed matter physics such as melting, freezing and glass transitions. The solidification of colloidal suspensions has long been a puzzling phenomenon with many unexplained features. Here we demonstrate and rationalize the existence of instability and metastability domains in cellular solidification of colloidal suspensions, by direct in situ high-resolution X-ray radiography and tomography observations. We explain such interface instabilities by a partial Brownian diffusion of the particles leading to constitutional supercooling situations. Processing under unstable conditions leads to localized and global kinetic instabilities of the solid/liquid interface, affecting the crystals morphology and particle redistribution behaviour.

---

[*] Correspondence should be addressed to S.D.: sylvain.deville@saint-gobain.com

Investigations of a solid/liquid interface stability have direct implications and practical applications in fields as diverse as metallurgy and materials science[1,2], crystal growth[3], cryopreservation[4], microfluidics[5], chromatography[6], food engineering[7], earth science[8] or biology[9]. In materials science, the cellular solidification of colloidal suspensions is becoming an attractive processing route for porous ceramic[10,11], polymeric[1] and metallic materials[12] or composites[13], with applications from tissue engineering to gas sensors[14] and catalysis[15] and where materials integrity and homogeneity is a requirement. Understanding the critical parameters controlling the stability of solidification interfaces in colloidal systems is a necessary step, its absence currently preventing a rationale control of the materials processing conditions. Already published investigations [16-18] of the solidification of colloidal suspensions are limited to the case of planar interface moving at low velocities (0.1-1 µm/s) by optical observations, where enough time is given to the system to reach equilibrium by Brownian diffusion. Typical velocities used when processing materials through solidification of colloidal suspensions are greater by at least one order of magnitude (10-100 µm/s). The system in these conditions is likely to be out of equilibrium.

What we understand so far of the solidification of colloidal suspensions is derived primarily from the analogies with dilute alloys systems, or the investigated behaviour of single particles in front of a moving interface and is still a subject of intense work. A more realistic, multi-particles model should account for the particles movement, the various possible interactions between the particles and the multiple interactions between the particles and the solid/liquid cellular interface. We can tackle the problem using the "colloids as big atoms" approach, applying Brownian diffusion mechanisms and Fick's law. Several specificities of colloidal systems must nevertheless be accounted for. Unlike alloys, colloidal suspensions exhibit various non-linear physical properties dependence with the concentration of particles, such as the freezing temperature curve, a variation of the segregation coefficient with the interface velocity and a dependence of particle diffusivity upon concentration[17]. The equilibrium situation with a planar interface at low velocity has recently been greatly clarified, both experimentally and theoretically[16-18]. Attempts towards modelling the system out of equilibrium with a non-planar interface are extremely complex. Direct experimental observations on which further theoretical analyses could be built and validated are still lacking, preventing further progress in this field.

In order to bring new experimental observations, we chose to investigate the stability of a cellular interface during directional solidification of colloidal suspensions by using X-ray radiography and tomography, under conditions typically used in processing routes. X-ray radiography and tomography experiments provide several advantages in this context [19] : transparent materials, like those used in reference [20], are no longer required since opaque materials can be imaged using the X-rays absorption contrast; a high spatial resolution (submicronic) can be obtained on synchrotron instruments, so that in some cases individual particles can be resolved, and a three-dimensional reconstruction of the particle's arrangement and the solvent crystals —ice crystals in this case— after complete solidification is possible.

In the cellular regime we investigate here, ice crystals are nucleating and growing into the suspension, eventually reaching a regime where they exhibit a cellular morphology. Particles are repelled by the moving cellular interface, and concentrate laterally between the growing crystals. The particle concentration in these intercrystals regions (referred to as the freezing zone in the paper) is increasing until the osmotic pressure exceeds the capillary pressure; the solid/liquid interface invades the interparticle space at a breakthrough concentration $C_{max}$ –typically 0.55-0.58– usually close to the maximum particle random packing[21]. The solidified body (or frozen zone) can therefore be divided into two domains: particle-free regions, corresponding to the cellular ice crystals free of particles, and particle-rich regions, corresponding to the concentrated particles entrapped in ice. This templating effect by the ice crystals is the structuring mechanism used to control the morphology of materials processed by directional solidification[11], yielding materials with unprecedented properties, as we have shown in the past[10,13]. We can differentiate these two types of domains using X-ray tomography images, the absorption of the beam in such regions being very different. The continuous cellular morphology along the solidification direction results in the observation of vertical absorption contrast bands in X-ray radiography, as seen in figure 1a. Since the absorption signal is averaged through the whole thickness of the sample in radiography, the pictures cannot provide a direct view of the crystal structure. The propagation of the interface can nevertheless be clearly observed, as well as the region close to the crystals tips, as shown later. We can use the absorption contrast obtained by tomography after a proper image treatment and analysis to quantify the relative particle-free and particle-rich phase fraction and to obtain three-dimensional representation of the crystals morphology in the frozen body.

Our measurements show kinetic and structural evidences of global instabilities of the displacement of the interface during solidification experiments at moderate velocities (figure 1, movie S1). The series of radiographs such as the one shown in figure 1a can be used to measure the evolution of the position of the interface with time. This, in turn, can be used to calculate the displacement velocity of the interface. The macroscopic contraction of the cold finger supporting the mold and the suspension, resulting from the temperature change during solidification, is negligible. These global instabilities are first revealed by sudden jumps of the interface (movie S1) corresponding to a local increase of the interface velocity (figure 1c), and resulting in bright bands parallel to the interface on the radiograph observations during solidification (figure 1a) associated with a local decrease of the particle-rich phase fraction (figure 1b).

The whole interface is affected at the same time by the global instability: bright bands are spanning the entire cross-section at a given location along the solidification direction in figure 1a. Limited edge effects, due to the difference of thermal conductivities between the suspension and the mold, induce a slight macroscopic curvature of the interface; for clarity, this is better illustrated in the magnified image of figure S1. We can describe the impact of the global instabilities on the crystals structure more precisely through the tomography reconstruction of the frozen body. Three-dimensional reconstruction after the tomography scans reveals that the ice crystals exhibit the same morphology before and after the instability with a continuity of the ice crystals along the solidification direction (figure 1d). The instability can be understood as inducing a growth of the ice crystals in the plane perpendicular to the solidification direction, which in turn induces a local decrease of the entrapped particles fraction, due to particles being rejected below and above the crystals growing transversally. Such transverse crystals can clearly be observed in figure 1d (black arrows). When we solidify the same suspension at higher velocities or using larger particles, instabilities are no longer observed, the obtained frozen structure is homogeneous and continuous along the solidification direction (figure 1e). These global instabilities are different from the local and already observed side branch instabilities leading to tip splitting or neighbouring primary dendrites (figure 1f). Localized instabilities, appearing only between the crystals, can also be observed at the same time, resulting in transverse crystals (white arrows in figure 1d), and inducing the creation of a large number of defects in the materials after sublimation of the ice crystals (figure S2). Such localized instabilities cannot be observed with the radiography observations, due to their local nature and averaging of the signal through the sample thickness. Repeating the solidification experiments with different interfacial velocities and particle sizes, we can assess the conditions under which the global instabilities appear. We can build a stability diagram with two domains: a metastability domain (green) where no instabilities are observed and therefore homogeneous crystals are obtained, corresponding to high interface velocity and/or large particles, and an instability domain (red) where global instabilities occur, for small particle size and/or low interface velocities (figure 2a). The reason why we refer to this situation as metastable and not as stable is discussed later in the paper.

We propose that these instabilities are due to partial diffusion of the particles ahead of the solidifying interface, leading to constitutional supercooling situations. The freezing point depression dependency on particles content has been reported previously[17] and is, unlike in the case of metallic alloys, non-linear. For convenience, these data have been replotted in figure S3. Although the plot shown here in figure S3 has been calculated and measured for bentonite particles, we can assume that the trend is similar for the alumina particles used in this study, the calculations being based on morphological parameters such as the particles hydrodynamic radius. The configuration of the system, with the definition of the various interfaces: the frozen zone/freezing zone interface and the freezing zone/suspension interface –although the latter one does not correspond to a physical interface– is shown in figure 3a. The corresponding experimental observation for metastable solidification is given in figure 3b. Under metastable conditions, where no instabilities occur, it is clear that the concentration of particles is restricted to the freezing zone. The diffusivity of the particles can be estimated using the Stokes-Einstein diffusivity of an isolated colloidal particle, corrected by the dimensionless diffusivity which accounts for the particle-particle interactions effects on the diffusivity[18]. The diffusivity can thus be expressed by:

$$D(R) = D_0(R)\widehat{D} \text{ (equation 1)}$$

where the dimensionless diffusivity $\widehat{D} = 0.502$ (see reference 18 for the details), and

$$D(R) = \frac{k_B T}{6\pi R \mu} \text{ (equation 2)}$$

is the Stokes-Einstein diffusivity of an isolated colloidal particle (R is the particle radius). The dimensionless diffusivity $\hat{D}$ can be calculated using the following parameters: $\phi=0.28$ (nominal fraction of particles in suspension), $\phi_p=0.64$ (monodispersed spheres, random packing), dynamic viscosity of water $\mu=1.787.10^{-3}$ Ns/m$^2$, T=273K. The actual dimensionless diffusivity is likely to be slightly lower, although the concentration dependence for particles of 0.1 µm and larger is low for a particle fraction below 0.55, see for example figure 3 of reference [18].

We can construct the expected particle concentration profile in the liquid phase along the position of the interface when the interface velocity is high (figure 4a), which corresponds to the experimental observations of figure 3b. The particle concentration in the particle-rich region of the solidified body is the breakthrough concentration $C_{max}$ defined previously. The initial concentration $C_{ini}$ is given by the formulation of the suspension (it has been verified that no segregation of the particles occurs within the time frame of the experiments). If the lateral growth velocity of the crystals is independent of the distance to the crystal tip, the shape of the growing crystals should be triangular, as sketched in figure 3a. Then, assuming that no Brownian diffusion can take place at high interface velocity, the particle concentration in the freezing zone should vary linearly from $C_{ini}$ up to a concentration where the interparticle interactions make further concentration more difficult; the particle concentration then progressively reaches $C_{max}$, deviating from the linear dependency. Using the freezing point depression of fig S3, we can construct the actual freezing point profile for this situation, along with the temperature profile. The discontinuity at the fr/fz interface is due to the latent heat (fig. 4b). The relative position of the actual temperature profile and the freezing point cannot be a priori determined. We make the hypothesis here that the temperature gradient is large enough that the actual temperature is lower than the freezing temperature.

Page 5/18

From the construction of figure 4b, the freezing zone appears to be in a constitutionally supercooled state, with the actual temperature below the solidification temperature. The zone ahead of the crystals tip is in a stable state. For high interface velocity, nucleation and growth cannot occur, so that the solidification occurs regularly with no instabilities, yielding homogeneous materials. This situation is therefore referred to as metastable, and corresponds to the metastable state zone defined in figure 2a.

The unstable situation occurs when enough time is given to the system for diffusion to occur, redistributing particles away from the concentrated freezing zone. The modified concentration and freezing point profile are given in figure 4c-d: the supercooled zone now extends beyond the crystals tips. We can verify qualitatively the concentration profile in this configuration experimentally, by plotting the grey level, directly dependent on the particle concentration, across the interface and within the freezing zone (figure 5). Note that the location of the line profile was selected to be located away from a pure ice crystal so that it is directly linked to the concentration of particles in the liquid phase. The observed particle concentration profile is qualitatively in excellent agreement with that expected from the model proposed here, providing a very good validation of the theory:

- the linear section of the particle concentration profile in between the crystals tips indicates that the concentration of the particles is induced by the lateral growth of the ice crystals exhibiting a triangular shape,
- the deviation from the linear trend when the particle concentration is high indicates that the particles interactions become non negligible and make further concentration difficult when approaching the fr/fz limit,
- the concentration profile with particle concentration greater than $C_{ini}$, extending beyond the crystals tips, indicates that the diffusion is partly occurring, redistributing particles away from the concentrated zone between the crystals tips.

One intriguing result is the dependency on particle size of the velocity leading to a transition between the stability domains (dashed area in figure 2a). We can expect the metastable to unstable transition to depend —at least— on two mechanisms: the development of a constitutionally supercooled zone ahead of the crystals tips, which provides the necessary conditions for the instabilities to appear, and the nucleation of the crystals in the supercooled zone. The first mechanism is directly related to the particle diffusivity (see equations 1 and 2), the dependency on particle size is therefore straightforward. The mechanisms controlling nucleation in such a complex system are more difficult to assess. Particles can probably act as preferential nucleation sites in the constitutionally supercooled zones, so that some influence of the particle size might also be expected: smaller particles mean more nucleation sites for an equivalent particle content. Assuming that the nucleation sites are related to the particle surface available at the crystal's tips for nucleation, a power-law behaviour can also be expected, although with an exponent 2 as the surface available for nucleation increases with the square of the particle size for a constant particle content. This is nevertheless still speculation only. An additional effect should be accounted for: the diffusion of the particles is somehow linked to the interface velocity, since the enrichment of particles in the freezing zone is due to the lateral growth of ice crystals. Faster growth kinetics will lead to faster enrichment of particles and therefore favour the diffusion away from the freezing zone. The particle size dependency of the transition between the stability regimes is still mysterious, but we can certainly expect the particle diffusivity to play a crucial role. Which mechanism is dominant and rate limiting is still to be assessed and deserves further exploration in the future.

We can now explain the development of the localized and global instabilities (figure 6). When the constitutionally supercooled zone extends beyond the tips of the crystals, new crystals are able to nucleate homogeneously in the supercooled zones. The strong anisotropy of growth kinetics of ice leads to a rapid growth of the favourably oriented crystals, leading to large crystals in the constitutionally supercooled zone, and similar to that observed for low interface velocity (see Fig. 1c of reference [18]). In the meantime, the cellular crystals entering the constitutionally supercooled zone should exhibit a much faster growth in the region $z_{fz/diff}$-$z_{diff/s}$ until the equilibrium is reached again (the actual temperature matches the solidification temperature, $z_{fz/diff} = z_{diff/s}$) when the particle concentration at the crystals tips corresponds to the nominal particle concentration in the suspension. This explains both the continuity of the cellular crystals in the solidified body (figure 1d) and the acceleration of the interface observed experimentally (figure 1c). Since the acceleration of the interface velocity corresponds to an unstable situation, the actual growth velocity is probably much greater than experimentally measured; the measure is currently limited by the radiography acquisition frequency used in these experiments (3 Hz). Furthermore, this mechanism accounts for the experimental observation that localized instabilities occur more often than global instabilities, comparing for instance figure 1a and S2, the periodicity of the global and localized instabilities being respectively a few hundreds of micrometers and a few micrometers. The nucleation and growth time available is more important in the freezing zone than ahead of the crystals tip. This should imply that a limited zone with only localized instabilities should exist in the stability diagram shown in figure 2a; this zone is represented by the dashed area between the metastable and unstable regions and its existence should be verified experimentally.

We can now plot a general stability diagram for the solidification of colloidal suspensions as a function of particle size and interface velocity (figure 2b), putting these results in a larger context. Instabilities of both type, affecting the integrity and homogeneity of the materials, are a major drawback of these processing routes, which must certainly be avoided. The proposed stability diagram is expected to be of great practical importance for the people processing materials through directional solidification. Using powders in the nanometre range, which can be required for materials with high surface area such as for catalysis applications, obtaining homogeneous and defect-free structures will be extremely difficult. The stability of the system being related to the diffusivity of the particles, workaround solutions might nevertheless be found: a modification of the suspension characteristics such as the viscosity can be used to lower the particles diffusivity. Since the transition between the stability domains is likely to be —at least partly, as discussed previously— determined by the particle diffusivity, this implies that a different stability diagram will be obtained when the formulation of the colloidal suspension is modified. Additives used to modify the crystals growth morphology, for example[22], will certainly affect the stability domain boundaries. To be of practical importance, the stability diagram will have to be assessed for each formulation.

The recent identification of a stable to unstable morphology transition[17] of the planar solidification interface in colloidal suspensions provided an accessible steady state to investigate the growth of morphological instabilities, which have numerous implications beyond materials processing. With the identified metastability region and corresponding metastable to unstable transition revealed here, it is not only possible to control the structure and the integrity of the resulting materials, but also to investigate the growth of morphological instabilities originating from a metastable state in a cellular regime, which could have implications not only in materials synthesis but also in various areas such as the design of microstructures in cast metal-matrix composites, organic electronics, microfluidics, geology, food engineering, biomaterials or cryobiology.

**Methods**

Slurries were prepared by mixing distilled water with a small amount (1 wt% of the powder) of ammonium polymethacrylate anionic dispersant (Darvan C, R. T. Vanderbilt Co., Norwalk, CT) and the alumina powder (see figure S4 and table S1). Slurries were ball-milled for 20 hrs with alumina balls and de-aired by stirring in a vacuum desiccator, until complete removal of air bubbles (typically 30 min).

Freezing experiments were performed by pouring the suspensions into a polypropylene mold (invisible to X-rays) placed in a cooling cell cooled using liquid nitrogen. The freezing kinetics were controlled by a heater placed at the bottom of the copper rod and a thermocouple also placed at the bottom. The experimental setup allows only linear ramp of temperature during cooling. Various cooling rates were used in the experiments, to investigate a wide range of interface velocities.

The experimental part of this study focuses on the use of a three-dimensional non-destructive technique, i.e. the high-resolution X-ray absorption radiography and tomography, to directly image nucleation and growth of the crystals in the suspension and the corresponding particles redistribution. Radiography is used to provide a dynamic visualisation of the solidification of the suspension. Tomography combines information from a large number of X-ray radiographs taken with different viewing angles of the sample, which should not move during the scan. The technique includes a computed step, i.e. a recalculation step during which a 3D map of the local absorption coefficients in each elementary volume of the sample is retrieved from the set of absorption radiographs. In this study, the HST program available at the ESRF was used [23]. The reconstructed map gives an indirect image of the microstructure. In the present study, the samples were scanned using a high-resolution X-ray tomograph located at the ESRF (beam line ID 19) in Grenoble (France). X-ray tomography was performed at a voxel size of $(1.4~\mu m)^3$. The energy was set to 20.5 keV. The distance between the sample and the detector was 20 mm. Because of the extremely high coherence of the X-ray beam on this beam line, absorption is not the only source of contrast in the obtained radiographs and phase contrast is also present, but in a small amount. A set of 1200 projections was taken within 180°. The detector was a CCD camera with 2048 × 2048 sensitive elements coupled with an X-ray-sensitive laser screen. Dynamics of solidification was followed by X-ray radiography, with an acquisition frequency of 3 Hz, which allows to precisely follow the interface evolution in two dimensions. The frozen structures after complete solidification were characterized in three dimensions afterwards using a low speed high-resolution tomography acquisition, and the results were then combined with the data obtained by radiography. To improve the acquisition frequency, the radiographs were acquired in a binning configuration (i.e. with a 1024x1024 pixels resolution, half that of tomography acquisitions).

Frozen samples were freeze-dried (Freeze dryer 8, Labconco, Kansas City, MI) for 24 hours. The green bodies thus produced were sintered in air for 2 hours at 1500°C, with heating and cooling rates of 5°C/min (1216BL, CM Furnaces Inc., Bloomfield, NJ). The microstructure of the samples was analyzed by scanning electron microscopy (ESEM, S-4300SE/N, Hitachi, Pleasanton, CA).

## Acknowledgements

We acknowledge the European Synchrotron Radiation Facility for provision of synchrotron radiation beam time and we would like to thank Elodie Boller and Jean-Paul Valade for their irreplaceable assistance in using beamline ID19. Financial support was provided by the National Research Agency (ANR), project NACRE in the non-thematic BLANC programme, reference BLAN07-2_192446.

## Authors contributions

S.D. and Ch.G. designed the research project, S.D. and E.M. designed the experiments, S.D., E.M., A.L., J.L., Ca.G. and A.B. performed the experiments, S.D. and G.B-G. analyzed the data, S.D., E.M. and G.B-G. wrote the paper. All authors discussed the results and implications and commented on the manuscript at all stages.

## Additional information

Supplementary information accompanies this paper on www.nature.com/naturematerials.

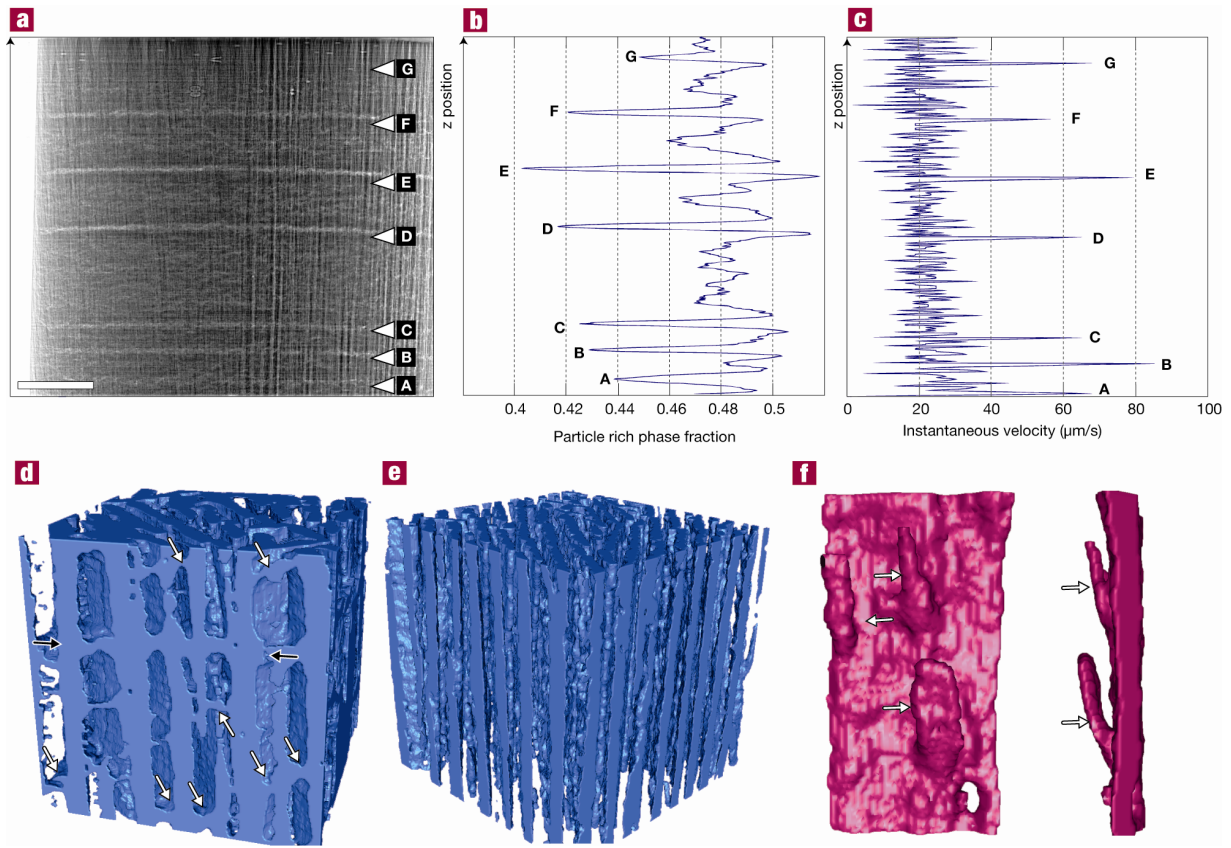

**Figure 1: Experimental observations of the development of global interface instabilities and their impact on the crystals structure, 0.2 µm particles. a**, radiography evidences of global instabilities, appearing as bright bands (less absorption of the beam). Interface displacement direction: bottom to top. Scale bar: 500 µm. **b,** particle-rich phase fraction in the frozen body. The particle-rich phase fraction is decreasing at locations corresponding to the development of the instabilities, **c,** instantaneous velocity of the interface. The sequence of radiographs acquired, as the interface is moving, can be used to track the interface position and provides a local velocity measurement. Spikes in the instantaneous velocity of the interface are observed, at locations corresponding to just before the bright bands and the decrease of particles fraction. The bright bands observed in **a**, corresponding to the instabilities of the interface velocities identified in **c**, also corresponds to zones of the sample with less absorption; the particles content is therefore lower. **d-e** 3D local observation by X-Ray tomography of the ice crystals, revealing lateral growth of ice crystals resulting from localized (white arrows) and global (black arrows) instabilities for 0.2 µm particles suspensions under unstable conditions (d), and their absence for 1.3 µm particles suspensions under metastable conditions (e). Blocks dimensions: 360x360x360 µm³. (f) neighbouring primary dendrites (arrows) resulting from side branch instabilities, front view (left) and side view (right), block dimensions: 30x70x140 µm³. The absence of a bright band in the radiograph **1a** at point G is due to the dynamics of instabilities

formation. A few seconds are necessary for evidence of instability to appear on the radiograph. The micrograph shown here was taken just a few seconds after the interface passed point G, the solid/liquid interface can be seen in the upper right part of the micrograph. The lateral crystals growth and particles redistribution resulting from the instability is not completed yet. The corresponding particle rich fraction shown in fig. 1b was measured after complete solidification of the sample. The spike in instantaneous velocity corresponding to point G and measured in fig. 1c can be clearly observed in movie S1.

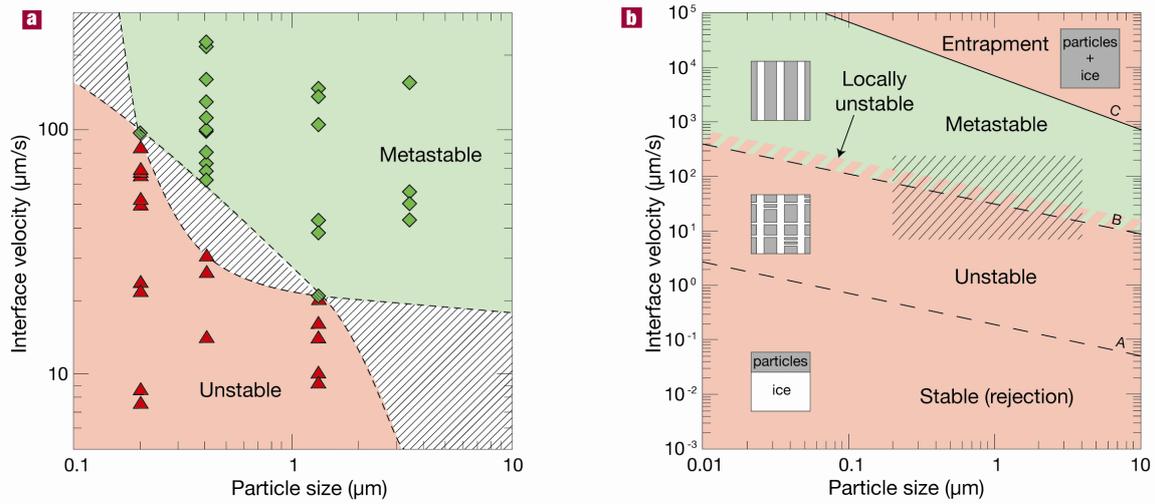

**Figure 2: Experimental conditions for the development of global instabilities: ice crystals stability and structure diagrams. a,** Metastability diagram as function of particle size and interface velocity, for a given particles fraction, based on various interface velocities and particle size conditions investigated in this paper. The dashed data points correspond to experiments where no instabilities were observed, although the measurement time span might be too short to be significant considering the chaotic frequency of instabilities. The dashed area is an estimation of the domain in which the metastable to unstable regime transition can occur; further experiments are necessary to situate more precisely the transition. **b,** General stability and structure diagram for the solidification of colloidal suspensions. The dashed area indicates the conditions investigated in this paper. At low velocities, the solid/liquid interface is planar and all particles are rejected. The interface becomes unstable when the velocity increases, due to constitutional supercooling situations created by particles Brownian diffusion ahead of the solidification front[16]. Above the critical velocities, all particles are entrapped[24]. The green areas indicate the metastable conditions suitable for processing homogeneous and defects-free cellular materials.

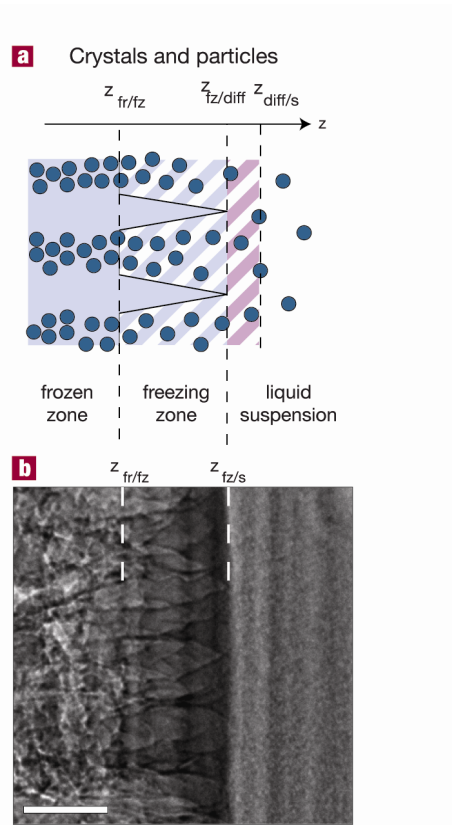

**Figure 3: Particles and crystals arrangement and definition of the interfaces location close to the crystals tips**. **a**, schematic representation. Acronyms for the position definition: fr: frozen zone, fz: freezing zone, diff: diffusion layer. **b**, high resolution X-ray radiography of the particle and crystals arrangement near crystals tips in metastable solidification conditions (interface displacement direction: left to right). The morphology of the crystals tip, exhibiting triangular shapes, is in good agreement with the model shown in figure 3a, supporting the hypothesis of a constant lateral growth kinetic (perpendicular to the solidification direction). The concentration of particles is restricted to the freezing zone, no diffuse layer ahead of the crystals tips can be observed. Scale bar: 50 µm.

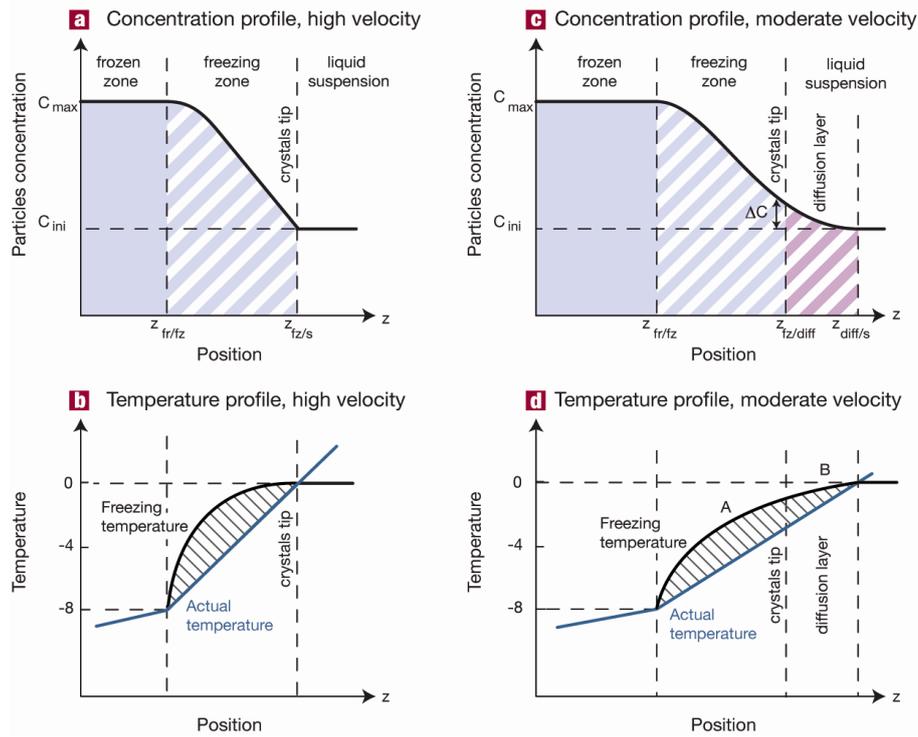

**Figure 4: Proposed mechanism for the development of instabilities: particle concentration, temperature and freezing point profile, for metastable and unstable situations**. **a,** Particle concentration profiles for high velocity (metastable situation). **b**, corresponding temperature and freezing point profiles. The dashed area indicates the zone of constitutional supercooling; global instabilities cannot appear ahead of the ice crystals tips. The development of localized instabilities in the freezing zone is limited by the available time for nucleation and growth of crystals in the constitutionally supercooled zone. **c**, particle concentration profiles for moderate velocity (unstable situation), where diffusion can occur. An increase of particle concentration ΔC is observed at the crystals tip position **d**, corresponding temperature and freezing point profile. The dashed area indicates the zone of constitutional supercooling, which now extends beyond the dendrites tips, allowing for global instabilities.

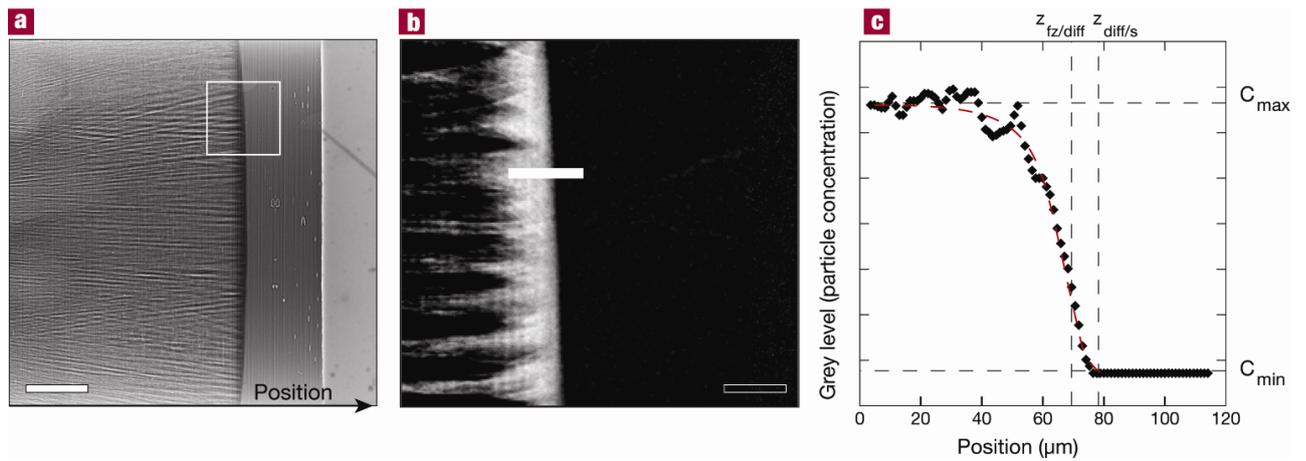

**Figure 5**: **Experimental observations of particle concentration profile under unstable conditions** (0.4 µm particles, interface displacement direction: left to right): **a,** Radiograph of the sample, **b,** detail from the panel a, with the background removed and grey level scale inverted. **c,** corresponding concentration profile (grey level) across the interface (white line in panel b). The crystals tip position defining the freezing zone/diffusion layer interface can be estimated, although the radiograph is the average signal from the whole sample thickness. The concentration profile is in good agreement with that expected from the proposed mechanism (figure 4c). Scale bars: a 500 µm, b 100 µm. The periodic ghost pattern (vertical stripes) visible in panels a and b is an acquisition artefact of the detector formed when subtracting the background, and is inducing the grey level periodic fluctuations observed in the freezing zone. The dashed line corresponds to the experimental data corrected to account for this ghost pattern.

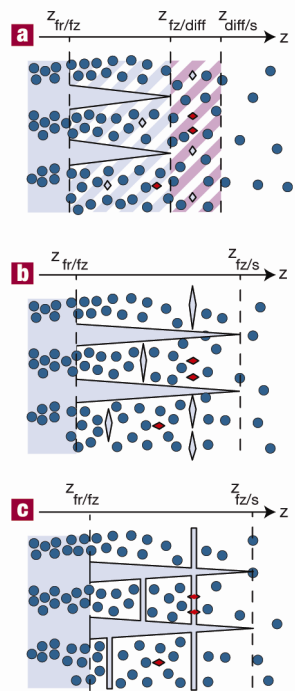

**Figure 6: Schematic sequence of the development of the instabilities**, with nucleation and growth of crystals in the constitutionally supercooled zones A (freezing zone) and B (diffusion layer ahead of the crystals tip). Favourably oriented nuclei (blue) grow faster than the unfavourably oriented ones (red). The rapid growth kinetics of the crystals in the constitutionally supercooled zone leads to crystals spanning transversely the entire freezing zone, as observed in figure 1d (black arrows).